\documentclass{iopjournal}
\usepackage{amsmath,amssymb,amsfonts}
\usepackage{bm, latexsym}
\usepackage{bbm}
\usepackage[normalem]{ulem}
\usepackage{cite}
\usepackage{mathrsfs}
\usepackage{mathtools}

\begin{document}

\articletype{Paper}

\newcommand{\cred}[1]{\textcolor{black}{#1}}

\title{Anomalous waiting-time distributions in postselection-free quantum many-body dynamics under continuous monitoring}



\author{Kazuki Yamamoto$^{1,2,3,4,*}$ and Ryusuke Hamazaki$^5$}

\affil{$^1$Research Institute for Innovation and Co-Creation, Osaka Metropolitan University, Sakai, Osaka 599-8531, Japan}

\affil{$^2$Department of Physics, Osaka Metropolitan University, Sumiyoshi, Osaka 558-8585, Japan}

\affil{$^3$Nambu Yoichiro Institute of Theoretical and Experimental Physics (NITEP), Osaka Metropolitan University, Sumiyoshi, Osaka 558-8585, Japan}

\affil{$^4$Department of Physics, Institute of Science Tokyo, Meguro, Tokyo 152-8551, Japan}

\affil{$^5$Nonequilibrium Quantum Statistical Mechanics RIKEN Hakubi Research Team, RIKEN Pioneering Research Institute (PRI), RIKEN iTHEMS, Wako, Saitama 351-0198, Japan}

\affil{$^*$Author to whom any correspondence should be addressed.}

\email{kazuki-yamamoto@omu.ac.jp}




\date{\today}

\begin{abstract}
We investigate waiting-time distributions (WTDs) of quantum jumps in continuously monitored quantum many-body systems, whose unconditional dynamics lead to the trivial infinite-temperature state. We demonstrate that the WTD of a half-chain subsystem exhibits an anomalous tail, markedly deviating from the Poissonian distribution in stark contrast to that of the whole system. By analyzing the spectral properties of the superoperator $\mathscr L_0$, which is defined by removing the jump terms associated with the half-chain subsystem from the full Liouvillian, we find that the long-time behavior with the anomalous tail of the half-chain WTD is governed by the eigenvalue $\lambda_0\:(<0)$ with the largest real part. We further reveal a qualitative change in the system-size dependence of $\lambda_0$ as a function of the measurement strength: for sufficiently weak measurement, $\lambda_0$ decreases proportionally to the system size, while for strong measurement, $\lambda_0$ scales independently of the system size, signaling the persistence of the anomalous half-chain WTD in the thermodynamic limit. The WTD is extracted solely from the spacetime record of quantum jumps $\{t_i,x_i\}$ and can be experimentally accessed without postselection. Our work establishes a spectral framework for understanding nontrivial WTDs in subsystems of monitored quantum dynamics and provides a novel diagnostics to assess many-body effects on WTDs.
\end{abstract}

\section{Introduction}
In recent years, the impact of measurements on quantum dynamics has attracted growing attention across a wide range of contexts, including nonunitary circuits and open many-body systems \cite{Muller12, Daley14, Ashida20, Harrington22, Fazio25, Hamazaki25rev, Roccati26}. A prominent example is the measurement-induced phase transition, where increasing the measurement strength drives a qualitative change in entanglement structures, such as a transition from volume-law to area-law entanglement scaling  \cite{Fisher23rev, Fisher18, Smith19, Skinner19, Yaodong19, Noel22, Koh22, Google23, Cao19, Alberton21, Turkeshi21, Turkeshi22, Piccitto22, Tang20, Fuji20, Szyniszewski20, Lunt20, Jian21, Van21, Doggen22, Minato21, Muller21, Buchhold21, Yamamoto23, Marcin23, Matsubara25, Mochizuki25, Matsubara26}. These discoveries have established measurements as an essential ingredient for engineering and exploring novel quantum phases beyond the unitary paradigm.
However, a major challenge in extracting physical observables from measurement-induced dynamics is the frequent reliance on postselection, despite efforts to evade the problem \cite{Gullans20L, Yaodong23, Altman24, Ippoliti21, Grover21, Yamamoto23, Mog23, Passarelli24, Max24, Skinner25, Barratt22L1, Agrawal24, Ippoliti24, Akhtar24, Hansveer25}. Since postselection typically incurs an exponential experimental overhead, it severely limits the experimental accessibility of many theoretically proposed quantities and obscures the direct characterization of measurement-induced many-body physics.

By contrast, the spacetime record of quantum jumps---namely, their occurrence times and positions $\{t_i,x_i\}$---is directly accessible in experiments without postselection \cite{Fazio25}. Such statistics of stochastic trajectories have been extensively studied from various perspectives, including dynamical activity and current fluctuations by using large-deviation statistics \cite{Garrahan18rev, Landi24, Garrahan07, Garrahan10, Raphael15, Barato15, Garrahan17, Carollo19, Horowitz20}. More recently, attention has turned to the role of many-body effects on quantum jump statistics \cite{Znidaric14, Hickey13, Lesanovski13, Carollo18A, Bao24, Cenap12, Buca14, Znidaric14B, Znidaric14E, Landi22, Matsumoto25}. It has been shown that the interplay between measurements and many-body effects can generate unusual phenomena in, e.g., spatiotemporal correlations \cite{Cech24, Cech25} and subsystem fluctuations \cite{Yamamoto25} in monitored quantum dynamics. These results highlight quantum jump statistics as a powerful and experimentally viable probe of measurement-induced many-body physics.

Among such statistical quantities, the distribution of waiting times between consecutive quantum jumps---known as the waiting-time distribution (WTD)---has a long history in probability theory and has been found to exhibit distinctive features in a wide range of physical systems \cite{Plenio98, Landi24}, including optics \cite{Cohen86, Byas88, Carmichael89, Bardou94}, electronic systems \cite{Albert12, Thomas13, Delteil14, Haack14, Dasenbrook15, Stegmann21, Schulz23}, and classical systems such as biological systems \cite{Skinner21}. Moreover, WTDs following the Lindblad dynamics are paradigmatic examples \cite{Landi24, Brandes08, Thomas14, Brandes16, Kosov16, Chia17, Landi21, Landi23, Coppola24, Radaelli24, Fresco24, Santos25, Sifft25, Soares25}, and many-body effects on them have been actively investigated, e.g., in quantum dots with source and drain \cite{Rajabi13, Ptaszynski17, Klein21}. However, little is known about the influence of many-body effects on WTDs in continuously monitored quantum systems whose unconditional steady state is the trivial infinite-temperature state \cite{Le24}. In fact, for particle-number measurements, it is known that the WTD of the entire system reduces to a trivial Poissonian distribution \cite{Landi24, Yamamoto25}. This naturally raises a fundamental question: can nontrivial measurement-induced many-body physics be encoded in the WTD, which is extracted solely from the postselection-free information $\{t_i,x_i\}$?

\begin{figure}[htbp]
\centering
\includegraphics[width=0.9\textwidth]{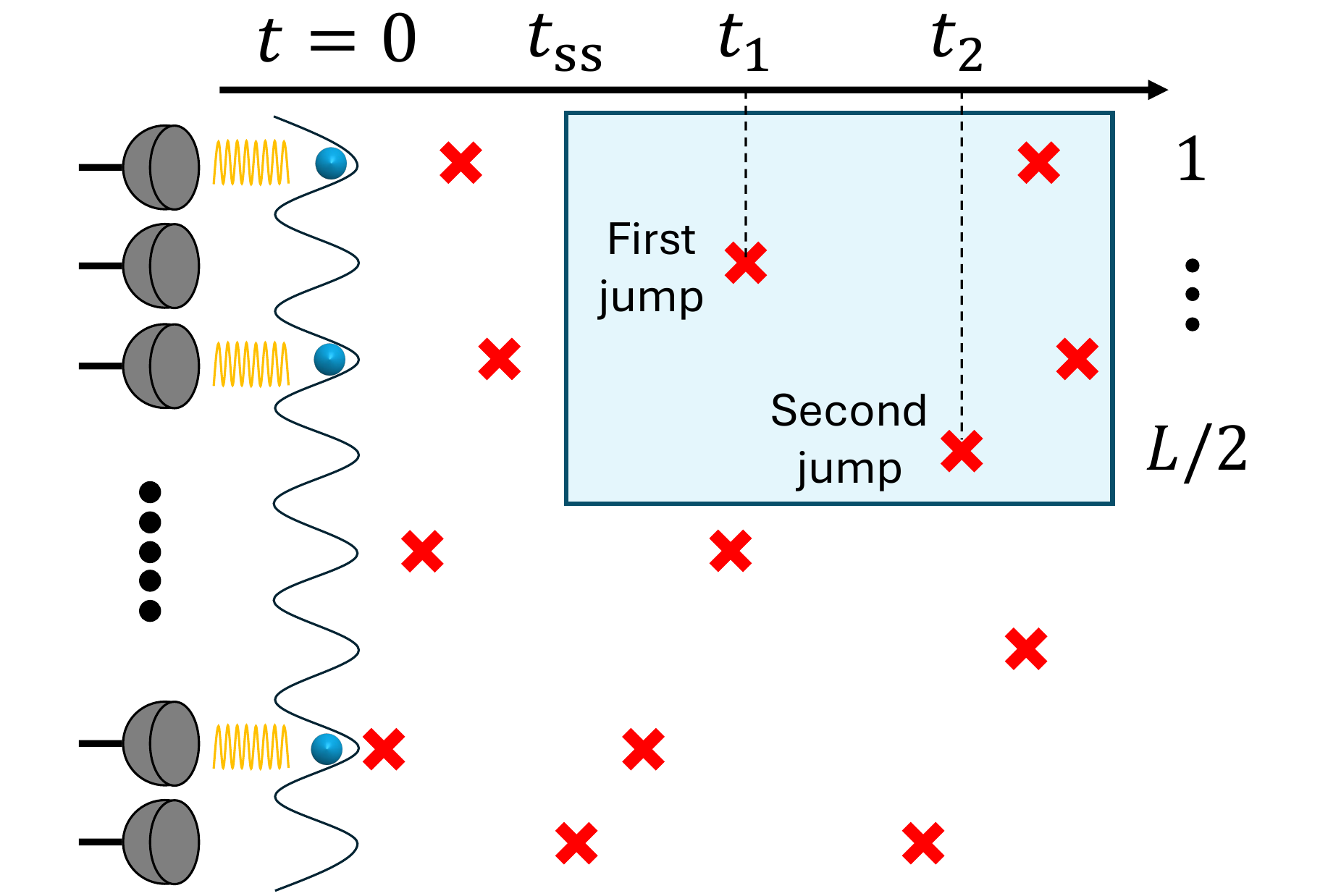}
\caption{Schematic figure of our setup. We focus on the first and the second jumps (red crosses) in a half chain after the system reaches the steady state (see text) and calculate the probability distribution of the waiting time $\tau\equiv t_2-t_1$ along trajectory realizations.}
\label{fig_schematic}
\end{figure}

In this work, we investigate half-chain WTDs of quantum jumps in the steady state of continuously monitored quantum many-body dynamics (see Fig.~\ref{fig_schematic} for the schematic figure). We show that the WTD of a half-chain subsystem $M\in[1, L/2]$ develops an anomalous tail, deviating significantly from the Poissonian distribution induced by measurements. We demonstrate that this anomalous behavior is fully characterized by the superoperator $\mathscr L_0$, defined by removing the jump terms associated with $M$ from the full Liouvillian. By systematically analyzing the spectral properties of $\mathscr L_0$, we find that, in contrast to the full Liouvillian, $\mathscr L_0$ does not possess a steady state (i.e., a zero eigenvalue), and the long-time behavior of the half-chain WTD is governed by the eigenvalue $\lambda_0\:(<0)$, which has the largest real part among all eigenvalues. We identify a qualitative change in the system-size scaling of $\lambda_0$ controlled by the measurement strength $\gamma$. In the weak-measurement regime, $\lambda_0$ decreases linearly with increasing the system size. On the other hand, in the strong-measurement regime, $\lambda_0$ does not depend on the system size, indicating that the anomalous tail remains robust in the thermodynamic limit. In summary, the long-time behavior of the half-chain WTD is described as
\begin{align}
W_{\mathrm{half}}(\tau) \sim e^{\lambda_0 \tau}
&=
\begin{cases}
e^{-\mathcal O(L)\tau}, \quad(\gamma \ll 1),\\
e^{-\mathcal O(1)\tau}, \quad [\gamma = \mathcal O(1)],
\end{cases}
\end{align}
which is numerically obtained by using the finite-size spectrum based on the exact diagonalization.
Since WTD can be extracted solely from the spacetime record of quantum jumps $\{t_i,x_i\}$, it constitutes an experimentally accessible observable that does not require postselection.

The remainder of this paper is organized as follows. In Sec.~\ref{sec_model}, we provide the model and the continuously monitored dynamics and obtain the WTD. Section~\ref{sec_L0} is devoted for the analysis of the superoperator $\mathscr L_0$, and we study half-chain WTDs in Sec.~\ref{sec_wtd}. A conclusion is given in Sec.~\ref{sec_conclusion}.

\section{Model and method}
In this section, we first explain the model and the dynamics under continuous monitoring. Then, we obtain the WTD in the steady state for a general subsystem $M$.
\label{sec_model}
\subsection{Quantum many-body dynamics under continuous monitoring}
We consider an interacting hard-core boson chain described by the Hamiltonian
\begin{align}
H=\sum_{i=1}^L\frac{J}{2}(b_{i+1}^\dag b_i + b_{i+1}b_i^\dag) + \sum_{i=1}^LJ_zn_{i+1}n_i,
\label{eq_HB}
\end{align}
where $b_i$ is the bosonic annihilation operator satisfying the hard-core constraint $b_i^2=0$, $n_i=b_i^\dag b_i$ is the particle number operator, and $L$ ($=2\mathbb N$) is the system size. We assume periodic boundary conditions $b_{L+1} = b_1$ to avoid edge effects in finite-size numerical simulations (note that the emergence of the anomalous WTD is not affected by the choice of boundary conditions) and set $J_z=J$ throughout this paper \cred{unless otherwise mentioned}. We note that, this model is equivalent to the Heisenberg model after the standard transformation, $b_i=S_i^-$, $b_i^\dag=S_i^+$, and $n_i=S_i^z+1/2$, where $S_i^\alpha\:(\alpha=+,-,z)$ are the spin-$1/2$ operators. We study the dynamics under continuous monitoring of a local particle number by employing the stochastic Schr\"odinger equation \cite{Fuji20, Yamamoto23, Yamamoto25}, given by
\begin{align}
d|\psi(t)\rangle=-iH |\psi(t)\rangle dt +\sum_{i=1}^L \bigg(\frac{n_i|\psi(t)\rangle}{\sqrt{\langle n_i\rangle}}-|\psi(t)\rangle \bigg)dN_i,
\label{eq_stochastic}
\end{align}
where $\langle\cdot\rangle$ denotes a quantum expectation value for the state $|\psi(t)\rangle$. Here, a discrete random variable $dN_i=0,1$ that counts the increment of a jump at site $i$ is chosen according to $dN_idN_j=\delta_{ij}dN_i$ and $E[dN_i]=\gamma\langle n_i\rangle dt$, where $E[\cdot]$ represents an ensemble average over the stochastic process, and $\gamma\:(>0)$ denotes the measurement strength. 
The total jump probability for the whole system is calculated as $p_\mathrm{tot}(t)=\gamma\sum_i \langle n_i\rangle dt=\gamma N dt$, which takes a constant value due to the particle number conservation $\sum_i \langle n_i\rangle=N$ under continuous measurement. This means that the WTD of the whole system is described by Poissonian statistics as discussed later.
Equivalently, Eq.~\eqref{eq_stochastic} is described as follows. Between jump events, the normalized state evolves as
\begin{align}
i\partial_t|\psi(t)\rangle=H|\psi(t)\rangle,
\end{align}
where the particle-number conservation is used. A jump occurs during the infinitesimal time interval $[t,t+dt]$ with probability $p_\mathrm{tot}(t)$,
and the jump channel $i$ is chosen with probability
\begin{align}
p_i(t)=\frac{\langle n_i\rangle}{N}.
\end{align}
Then, the state is updated as
\begin{align}
|\psi(t+dt)\rangle=\frac{n_i|\psi(t)\rangle}{\sqrt{\langle n_i\rangle}}.
\end{align}
In the following, we assume that the initial state is prepared in the N\'eel state $|1010\cdots\rangle$ at half filling.

Importantly, Eq.~\eqref{eq_stochastic} is obtained by unraveling the Lindblad equation \cite{Lindblad76} with the use of the quantum trajectory method \cite{Daley14, Yamamoto23}. The general form of the Lindblad equation is written as
\begin{align}
\dot{\rho}=\mathscr L (\rho)
&=-i[H,\rho]-\frac{1}{2}\sum_{i=1}^L(L_i^\dagger{L}_i\rho+\rho{L}_i^\dagger{L}_i-2L_i\rho{L}_i^\dagger),
\label{eq_Lindblad}
\end{align}
In this work, we define that the term ``Liouvillian" is used to describe the Lindblad dynamics that generates the completely positive and trace-preserving map \cite{Lindblad76, Daley14}. Also, to distinguish superoperators from operators expressed in the ladder representation, we explicitly denote superoperators using calligraphic fonts. By substituting the special jump operator $L_i=\sqrt \gamma n_i$ in our study, Eq.~\eqref{eq_Lindblad} is obtained from the ensemble average of Eq.~\eqref{eq_stochastic}. It should be noted that, under the particle number measurement, the state ${\rho}(t)$ averaged over the measurement outcomes becomes a maximally mixed state $\rho_\mathrm{ss}= {I}/D_0$ after long times irrespective of the measurement strength. Here, $I$ is the identity matrix and $D_0$ is the dimension of the Hilbert space in the half-filling sector.

\subsection{Waiting-time distributions for a subsystem $M$}
Here, we focus on the time evolution of a single quantum trajectory and consider the time interval $\tau$ between the first and the second quantum jumps after the time $t=t_{\mathrm{ss}}$, at which the ensemble-averaged density matrix sufficiently reaches its stationary state as $\rho(t_{\mathrm{ss}})\simeq \rho_{\mathrm{ss}}$. This time interval is referred to as the waiting time, and its probability distribution over stochastic trajectories is denoted as the WTD \cite{Landi24} (see Fig.~\ref{fig_schematic} for the schematic figure of the half-chain WTD). When we restrict our attention to quantum jumps occurring within a subsystem $M$, the corresponding superoperator $\mathscr L_0$, in which quantum jumps in the subsystem $M$ are excluded from the full Liouvillian $\mathscr L$, is defined as
\begin{align}
\mathscr L_0 \equiv \mathscr L - \sum_{i\in M}\mathscr L_i, 
\label{eq_nojump}
\end{align}
where $\mathscr L_i (\rho) \equiv L_i \rho L_i^\dagger$. Physically, this situation corresponds to cases in which quantum jumps outside the subsystem $M$ are inaccessible, or when one is interested solely in the jump statistics associated with the subsystem $M$.
We note that, in experiments, quantum jumps belonging to the subsystem should be interpreted as detection events associated with a selected set of monitoring channels \cite{Clemens03}.
Equation~\eqref{eq_nojump} thus describes the time evolution under the constraint that no quantum jumps occur in $M$, while we have no restrictions for the jumps in the complementary subsystem $\bar{M}$. When all jump operators are monitored, $\mathscr L_0$ reduces to the non-Hermitian Hamiltonian governing the no-jump evolution. In our study, quantum jumps are described by particle-number operators.

With this formulation, the WTD for the subsystem $M$ is computed using the superoperator $\mathscr L_0$ given in Eq.~\eqref{eq_nojump}. We first introduce the conditional stationary density matrix $\rho_c(t_{\mathrm{ss}})$ such that its ensemble average satisfies
$E[\rho_c(t_{\mathrm{ss}})]=\rho(t_{\mathrm{ss}})=\rho_{\mathrm{ss}}$.
For a quantum trajectory described by $\rho_c(t_{\mathrm{ss}})$, the probability that a quantum jump occurs for the first time at site $i$ after a time interval $t$ is given by
$p_{\mathrm{no},i}^c(t)\equiv\mathrm{Tr}[\mathscr L_i e^{\mathscr L_0 t}\rho_c(t_{\mathrm{ss}})]dt$. Then, after a quantum jump, the conditional density matrix is updated according to
\begin{align}
\rho^\prime_c(t)\equiv
\frac{\mathscr L_i e^{\mathscr L_0 t}\rho_c(t_{\mathrm{ss}})}
{\mathrm{Tr}[\mathscr L_i e^{\mathscr L_0 t}\rho_c(t_{\mathrm{ss}})]}
=
\frac{\mathscr L_i e^{\mathscr L_0 t}\rho_c(t_{\mathrm{ss}})\,dt}
{p_{\mathrm{no},i}^c(t)}.
\label{eq_update_i}
\end{align}
Since the subsequent time evolution until the next jump is governed by the superoperator $\mathscr L_0$, the probability that a quantum jump occurs at site $j$ after an additional waiting time $\tau$ with respect to the conditional density matrix \eqref{eq_update_i} is given by
\begin{align}
P_0&[dN_j(t+\tau)=1|dN_i(t)=1,\rho_c(t_{\mathrm{ss}})]\equiv
\mathrm{Tr}[\mathscr L_j e^{\mathscr L_0 \tau}\rho^\prime_c(t)]d\tau
=
\frac{\mathrm{Tr}[\mathscr L_j e^{\mathscr L_0 \tau}\mathscr L_i e^{\mathscr L_0 t}\rho_c(t_{\mathrm{ss}})]dt d\tau}
{p_{\mathrm{no},i}^c(t)}.
\label{eq_WTDcond}
\end{align}
Here, $P_0[\:\cdot\:|\:\cdot\:]$ denotes the conditional probability that, after time $t_{\mathrm{ss}}$, no quantum jumps occur within the monitored subsystem $M$ other than the two specified events. We note that, in calculating current fluctuations of quantum jump statistics \cite{Landi24, Yamamoto25}, no restriction is imposed between two subsequent jumps, and hence the factor $e^{\mathscr L_0 t}$ does not appear.

We are now in a position to calculate the WTD after the density matrix has reached the stationary state $\rho_{\mathrm{ss}}$. When generally focusing on quantum jumps within a subsystem $M$, the probability per unit time that the first jump occurs at site $i$ for some time and the second jump occurs at site $j$ after a waiting time $\tau$ is given by
\begin{align}
W(\tau,j; i)
&=
\int_0^\infty \frac{1}{d\tau}
E\!\left[
P_0[dN_j(t+\tau)=1|dN_i(t)=1,\rho_c(t_{\mathrm{ss}})]
p_{\mathrm{no},i}^c(t)
\right]\notag\\
&=
\int_0^\infty
\mathrm{Tr}[\mathscr L_j e^{\mathscr L_0 \tau}\mathscr L_i e^{\mathscr L_0 t}\rho_{\mathrm{ss}}]dt .
\label{eq_WTDij}
\end{align}
This quantity is nothing but the WTD.
We note that the following analysis of the WTD is given by using $\mathscr L_0$ for the ensemble-averaged dynamics and does not rely on the specific realization of the trajectory.

Importantly, the WTD satisfies the normalization condition, given by
\begin{align}
\sum_{i,j\in M}\int_0^\infty d\tau\, W(\tau, j; i)=1.
\label{eq_WTDnorm}
\end{align}
This is explicitly shown as follows. We can calculate the integrand in Eq.~\eqref{eq_WTDnorm} as
\begin{align}
\sum_{i,j\in M} W(\tau,j;i)=-\int_0^\infty \mathrm{Tr}[\mathscr L_0 e^{\mathscr L_0 \tau}\sum_{i\in M}\mathscr L_i e^{\mathscr L_0 t}\rho_{\mathrm{ss}}]dt
=-\frac{d}{d\tau}p^\prime_\mathrm{no}(\tau),
\end{align}
where we have used Eqs.~\eqref{eq_nojump} and \eqref{eq_WTDij} together with the fact that $\mathrm{Tr}[\mathscr L A]=0$ for any operator $A$. Here, we have introduced
\begin{align}
p^\prime_\mathrm{no}(\tau)\equiv\int_0^\infty \mathrm{Tr}[e^{\mathscr L_0 \tau}(\mathscr L - \mathscr L_0) e^{\mathscr L_0 t}\rho_{\mathrm{ss}}]dt.
\label{eq_WTDprob}
\end{align}
Then, we find
\begin{align}
\sum_{i,j\in M}\int_0^\infty d\tau\, W(\tau, j; i)=p^\prime_\mathrm{no}(0)-p^\prime_\mathrm{no}(\infty),
\label{eq_WTDijdetail}
\end{align}
where $p^\prime_\mathrm{no}(0)$ is evaluated as
\begin{align}
p^\prime_\mathrm{no}(0)=-\int_0^\infty \mathrm{Tr}[\mathscr L_0 e^{\mathscr L_0 t}\rho_{\mathrm{ss}}]dt
=-\int_0^\infty\left[\frac{d}{dt}p_\mathrm{no}(t)\right]dt
=p_\mathrm{no}(0)-p_\mathrm{no}(\infty),
\end{align}
with
\begin{align}
p_\mathrm{no}(t)\equiv\mathrm{Tr} [e^{\mathscr L_0 t}\rho_{\mathrm{ss}}].
\end{align}
By using $p_\mathrm{no}(0)=1$ and assuming that $p_\mathrm{no}(\infty)=0$, which indicates that the probability of observing no quantum jumps in the subsystem $M$ after infinitely long times is zero, we obtain $p^\prime_\mathrm{no}(0)=1$. Similarly, by assuming that $p^\prime_\mathrm{no}(\infty)=0$ in Eq.~\eqref{eq_WTDijdetail}, which implies that the second jump in Eq.~\eqref{eq_WTDprob} should occur in the subsystem $M$ at some time $0\le\tau\le\infty$, we arrive at Eq.~\eqref{eq_WTDnorm}.

\section{Superoperator $\mathscr L_0$}
\label{sec_L0}
To analyze the WTD, it is convenient to introduce the vectorization of the density matrix when computing physical quantities. We therefore summarize its definition and basic properties for the ladder representation of the superoperator $\mathscr L_0$. Since the properties of $\mathscr L_0$ are not widely discussed in the literature, it is important to present them in comparison with those of the standard Liouvillian.

\subsection{Ladder representation of $\mathscr L_0$}
We begin by defining the following equation for the density matrix associated with $\mathscr L_0$:
\begin{align}
\mathscr L_0 (\rho)=\mathscr L(\rho)-\sum_{i\in M}L_i\rho L_i^\dag,
\label{eq_Lindblad0}
\end{align}
where $\mathscr L$ denotes the conventional Liouvillian superoperator that satisfies Eq.~\eqref{eq_Lindblad}.
We introduce the right and left eigenmodes of $\mathscr L_0$ and consider its spectral decomposition. The right and left eigenvalue equations are defined as
\begin{gather}
\mathscr L_0 (\rho_\alpha^R)=\lambda_\alpha \rho_\alpha^R,\label{eq_right}\\
\tilde{\mathscr L_0} (\rho_\alpha^L)=\lambda_\alpha^* \rho_\alpha^L,\label{eq_left}
\end{gather}
where $\rho_\alpha^R$ and $\rho_\alpha^L$ represent the right and left eigenmodes of the density matrices, and $\lambda_\alpha$ is the corresponding eigenvalue of $\mathscr L_0$.
Here, the index $\alpha$ takes the values $\alpha=0,1,\cdots,D_0^2-1$, and $\tilde{\mathscr L}_0$ is defined through the relation
$\mathrm{Tr}[A\mathscr L_0(B)]=\mathrm{Tr}[\tilde{\mathscr L}_0(A)B]$, which implies
\begin{align}
\tilde{\mathscr L}_0 (A)=\tilde{\mathscr L}(A)-\sum_{i\in M}L_i^\dag A L_i,
\label{eq_Liouvillianad0}
\end{align}
where $\tilde{\mathscr L}$ denotes the adjoint Liouvillian \cite{Breur02}, defined by
\begin{align}
\mathscr {\tilde L}(A) = i[H, A] +\sum_i \left(L_i^\dag A L_i -\frac{1}{2}\left\{L_i^\dag L_i, A\right\}\right).
\label{eq_Liouvillianad}
\end{align}
By explicitly taking the Hermitian conjugate of both sides of Eqs.~\eqref{eq_Lindblad0} and \eqref{eq_Liouvillianad0}, and substituting them into the Hermitian conjugate of Eqs.~\eqref{eq_right} and \eqref{eq_left}, we find
\begin{gather}
\mathscr L_0 (\rho_\alpha^R)=\lambda_\alpha \rho_\alpha^R\xrightarrow{\dag}\mathscr L_0 (\rho_\alpha^{R\dag})=\lambda_\alpha^* \rho_\alpha^{R\dag},\label{eq_eigen1}\\
\tilde{\mathscr L}_0 (\rho_\alpha^L)=\lambda_\alpha^* \rho_\alpha^L\xrightarrow{\dag}\tilde{\mathscr L}_0 (\rho_\alpha^{L\dag})=\lambda_\alpha \rho_\alpha^{L\dag}.\label{eq_eigen2}
\end{gather}
This implies that the eigenvalues of $\mathscr L_0$ appear either as real numbers or as complex conjugate pairs.

Next, we consider a general density matrix
\begin{align}
\rho=\sum_{i,j=1}^{D_0} \rho_{ij}|i\rangle \langle j |,
\end{align}
and introduce a basis transformation $|i\rangle \langle j |\mapsto |ij)\coloneqq|i\rangle\otimes|j\rangle \in \mathcal H \otimes \mathcal H$,
which amounts to transposing operators acting on the bra vector that spans the Hilbert space $\mathcal H$. Under this transformation, the superoperator $\mathscr L_0$ given in Eq.~\eqref{eq_Lindblad0} is transformed into the tensor product of operators as
\begin{align}
\mathcal L_0=\mathcal L -\sum_{i\in M}L_i\otimes L_i^*.
\end{align}
Here, $\mathcal L$ denotes the Liouvillian in the ladder representation, given by
\begin{align}
\mathcal L&=-i(H\otimes I-I\otimes H^T)+\sum_i\left\{L_i\otimes L_i^*-\frac{1}{2}(L_i^\dag L_i\otimes I +I\otimes L_i^T L_i^*)\right\},\notag\\
&=-i(H_\mathrm{eff}\otimes I-I\otimes H_\mathrm{eff}^*)+\sum_i L_i\otimes L_i^*,
\end{align}
where $I$ stands for the identity operator acting on the Hilbert space $\mathcal H$. Note, for example, that matrix elements satisfy $\langle i |H_\mathrm{eff}^*|j\rangle =\langle i |H_\mathrm{eff} |j\rangle^*$.
Similarly, for $\tilde{\mathscr L}_0$, we obtain the ladder representation from Eq.~\eqref{eq_Liouvillianad0} as
\begin{align}
\tilde{\mathcal L}_0=\tilde {\mathcal L } -\sum_{i\in M}L_i^\dag \otimes L_i^T,
\end{align}
where $\tilde{\mathcal L}$ is the ladder representation of the adjoint Liouvillian, given by
\begin{align}
\tilde {\mathcal L}&=i(H\otimes I-I\otimes H^T)+\sum_i\left\{L_i^\dag \otimes L_i^T-\frac{1}{2}(L_i^\dag L_i\otimes I +I\otimes L_i^T L_i^*)\right\},\notag\\
&=i(H_\mathrm{eff}^\dag\otimes I-I\otimes H_\mathrm{eff}^T)+\sum_i L_i^\dag \otimes L_i^T.
\end{align}
Thus, we find that the important relation holds:
\begin{align}
\tilde {\mathcal L}_0 =\mathcal L^\dag_0.
\end{align}
Using these representations, the eigenvalue equations given in Eqs.~\eqref{eq_eigen1} and \eqref{eq_eigen2} can be rewritten as
\begin{gather}
\mathcal L_0 |\rho_\alpha^R)=\lambda_\alpha |\rho_\alpha^R),\\
\tilde{\mathcal L}_0 |\rho_\alpha^L)=\lambda_\alpha^* |\rho_\alpha^L).
\end{gather}
Here, the inner product, which is defined as $(A|B)=\mathrm{Tr}[A^\dag B]$ with $(A|={|A)}^\dag$, satisfies $(A|\mathcal L_0 B)=(\tilde{\mathcal L}_0A|B)$.

\subsection{Typical properties of the operator $\mathcal L_0$}
We obtain several important properties that the operator $\mathcal L_0$ satisfies. In this subsection, we assume that $\mathcal L_0$ is always diagonalizable and that the eigenmode hosting the largest real part is unique, so that eigenvalues are arranged according to
\begin{align}
0>\lambda_0>\mathrm{Re}[\lambda_1]\ge\mathrm{Re}[\lambda_2]\ge\cdots\ge\mathrm{Re}[\lambda_{D_0^2-1}],
\label{eq_eigenvalue}
\end{align}
(the fact $\lambda_0<0$ will be shown below). In addition, we consider the situation where the eigenstate $\rho_0^R$ fulfills the following condition
\begin{align}
\sum_{i\in M}\mathrm{Tr}[L_i^\dag L_i\rho_0^R]>0.
\label{eq_assumption1}
\end{align}
First of all, we remark that, for eigenvalues satisfying $\lambda_\alpha\neq\lambda_\beta$, the corresponding eigenmodes satisfy the biorthogonality
\begin{align}
(\rho_\alpha^L|\rho_\beta^R)=\delta_{\alpha\beta}.
\label{eq_biorthogonality}
\end{align}
Here, we have assumed that the eigenstates are normalized such that $(\rho_\alpha^L|\rho_\alpha^R)=1$. Then, $\mathcal L_0$ satisfies the following properties: the spectral decomposition
\begin{align}
\mathcal L_0 =\sum_{\alpha=0}^{D_0^2-1}\lambda_\alpha|\rho_\alpha^R)(\rho_\alpha^L|,
\label{eq_spectral}
\end{align}
and the completeness condition
\begin{align}
\sum_{\alpha=0}^{D_0^2-1}|\rho_\alpha^R)(\rho_\alpha^L|=I_{D_0^2},
\label{eq_complete}
\end{align}
where $I_{D_0^2}$ denotes the $D_0^2\times D_0^2$ identity matrix. It should be noted that, when $\mathcal L_0$ is diagonalizable, the biorthogonality \eqref{eq_biorthogonality}, the spectral decomposition \eqref{eq_spectral}, and the completeness condition \eqref{eq_complete} naturally holds. 
Using Eqs.~\eqref{eq_spectral} and \eqref{eq_complete}, the time-evolution operator is given by
\begin{align}
e^{\mathcal L_0 t}=\sum_{\alpha=0}^{D_0^2-1}e^{\lambda_\alpha t}|\rho_\alpha^R)(\rho_\alpha^L|.
\label{eq_timeevolution}
\end{align}
In contrast to the standard Liouvillian, $\lambda_0\neq0$ in general, and thus $\mathcal L_0$ does not possess an infinite-temperature steady state corresponding to the zero eigenvalue. In the case of the full Liouvillian, it is known that all the right eigenmodes except for the steady state satisfy $\mathrm{Tr}[\rho_\alpha^R]=0$, which is guaranteed by the trace preservation of the density matrix under the time evolution described by the Liouvillian. However, in the case of the superoperator $\mathscr L_0$, $\mathrm{Tr}[\rho_\alpha^R]\neq0$ holds for the right eigenmodes including $\rho_0^R$, when the time evolution $e^{\mathscr L_0 t}$ does not preserve the trace of the eigenmodes.

Next, we show that $\lambda_0<0$. Defining
$|\rho_0^R(t))=e^{\mathcal L_0 t}|\rho_0^R)$, we obtain
\begin{align}
\frac{d}{dt}(I|\rho_0^R(t))=\frac{d}{dt}(I|e^{\mathcal L_0 t}|\rho_0^R)=\lambda_0 e^{\lambda_0 t}\mathrm{Tr}[\rho_0^R].
\label{eq_trace}
\end{align}
At the same time, we focus on the fact that the left-hand side in Eq.~\eqref{eq_trace} can be rewritten using the definition \eqref{eq_Lindblad0} as
\begin{align}
\frac{d}{dt}(I|\rho_0^R(t))=\mathrm{Tr}[\mathscr L_0 \rho_0^R (t)]=-\sum_{i\in M} \mathrm{Tr}[L_i^\dag L_i \rho_0^R (t)]=-e^{\lambda_0 t}\sum_{i\in M} \mathrm{Tr}[L_i^\dag L_i \rho_0^R],
\label{eq_trace2}
\end{align}
where we have used $\mathrm{Tr}[\mathscr L \rho_0^R(t)]=0$ for the Liouvillian,
which immediately follows from the direct calculation by substituting Eq.~\eqref{eq_Lindblad}.
Using Eqs.~\eqref{eq_trace} and \eqref{eq_trace2}, we obtain
\begin{align}
\lambda_0 \mathrm{Tr} [\rho_0^R] = - \sum_{i\in M}\mathrm{Tr}[L_i^\dag L_i\rho_0^R].
\label{eq_positive}
\end{align}
Since $e^{\mathscr L_0 t}\rho(0)$ does not necessarily have a unit trace, physical states are given by $e^{\mathscr L_0t}\rho(0)/\mathrm{Tr}[e^{\mathscr L_0t}\rho(0)]$, where $\rho(0)$ denotes the initial density matrix.
At long times, the dynamics should be dominated by the eigenvalue $\lambda_0$ with the largest real part:
\begin{align}
\lim_{t\to\infty}\frac{e^{\mathscr L_0t}\rho(0)}{\mathrm{Tr}[e^{\mathscr L_0t}\rho(0)]}=\rho_0^R.
\label{eq_quasisteady}
\end{align}
Moreover, we can show that $e^{\mathscr L_0 t}$ is a completely positive map (see \ref{app_cp} for the proof). Therefore, from Eq.~\eqref{eq_quasisteady}, the right eigenmode $\rho_0^R$ is positive semidefinite. In addition, the right-hand side of Eq.~\eqref{eq_positive} is negative by assumption \eqref{eq_assumption1}, which holds when particles are not localized in the subsystem $\bar M$. Using these facts in Eq.~\eqref{eq_positive}, we conclude that $\lambda_0<0$.
In this sence, $\lambda_0$ is distinct from the spectral gap of the full trace-preserving Liouvillian. It characterizes the conditional no-detection dynamics in the long-time regime rather than the ordinary relaxation toward the stationary state.

\begin{figure}[tb]
\centering
\includegraphics[width=0.95\textwidth]{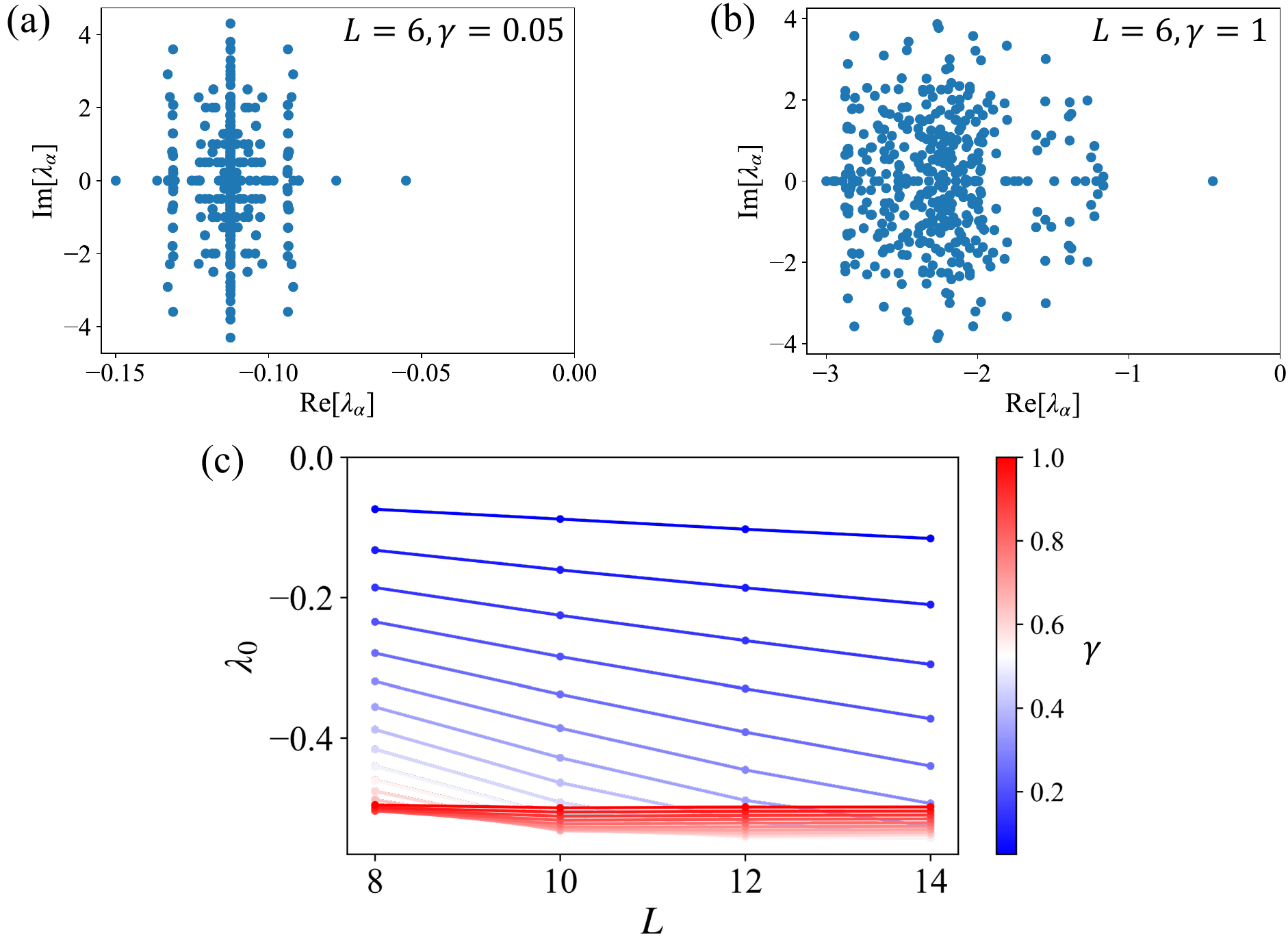}
\caption{Eigenspectrum of the superoperator $\mathscr L_0$ for the Heisenberg model under continuous monitoring for (a) $L=6$, $\gamma=0.05$ and (b) $L=6$, $\gamma=1$. The eigenvalue with the largest real part, $\lambda_0$, is negative in contrast to the case of the Liouvillian. (c) System-size dependence of $\lambda_0$, for different measurement strengths $\gamma$ as indicated by the colorbar. 
For weak measurement strength, $\lambda_0$ decreases proportionally to the system size, but for strong measurement, $\lambda_0$ is independent of the system size. 
}
\label{fig_nojumpop}
\end{figure}

Finally, to illustrate that the real parts of all eigenvalues of $\mathscr L_0$ are strictly negative, we plot the eigenspectrum of the superoperator $\mathscr L_0$ obtained by the exact diagonalization in Fig.~\ref{fig_nojumpop}. We find that $\lambda_0$, the right-most eigenvalue in Figs.~\ref{fig_nojumpop}(a) and \ref{fig_nojumpop}(b), is indeed negative and does not become zero. This indicates that, though the whole distribution of the eigenspectrum of $\mathscr L_0$ resembles that of the well-known Liouvillian (in the sense that the eigenvalues appear in complex conjugate pairs), $\mathscr L_0$ exhibits distinct properties from those of the Liouvillian. Also, for the weak measurement strength $\gamma=0.05$ shown in Fig.~\ref{fig_nojumpop}(a), the distribution separates into several clusters, whereas for the strong measurement strength $\gamma=1$ shown in Fig.~\ref{fig_nojumpop}(b), it leads to a broadly delocalized one. Furthermore, the system-size scaling of $\lambda_0$ is depicted in Fig.~\ref{fig_nojumpop}(c) for the measurement strength $\gamma=0.05, 0.1, ..., 1$. For small $\gamma$, $\lambda_0$ decreases proportionally to the system size $L$, whereas for large $\gamma$, $\lambda_0$ stays constant irrespective of the system size. 
\cred{Table \ref{tab_lambda0} shows the system-size dependence of $\lambda_0$ for strong measurement strengths.
We clearly see that $\lambda_0$ stays almost constant irrespective of the system size.}
As discussed in Sec.~\ref{sec_wtd} below, such a difference in the system-size dependence of $\lambda_0$ affects the persistence of the anomalous tail of the half-chain WTD.
We note that the system size used in the calculation is rather small to extract the critical value of $\gamma$ that characterizes the crossover.

\begin{table}[tb]
	\centering
	\caption{\cred{System-size dependence of $\lambda_0$ for strong measurement strengths $\gamma=1,2$, and $5$. The values of $\lambda_0$ stay almost constant irrespective of the system size.}}
	\label{tab_lambda0}
     \begin{tabular}{cccc} \hline \hline
    ~~~$L$~~~ & ~~~$\lambda_0\:(\gamma=1)$~~~ & ~~~$\lambda_0\:(\gamma=2)$~~~ & ~~~$\lambda_0\:(\gamma=5)$~~~\\ \hline
    $8$ & $-0.495$ & $-0.379$ & $-0.189$ \\
    $10$ & $-0.499$ & $-0.378$ & $-0.189$\\
    $12$ & $-0.498$ & $-0.378$ & $-0.189$\\
    $14$ & $-0.498$ & $-0.378$ & $-0.189$\\
    $16$ & $-0.498$ & $-0.378$ & $-0.189$\\ \hline
    \end{tabular}
\end{table}

\begin{figure}[tb]
\centering
\includegraphics[width=0.9\textwidth]{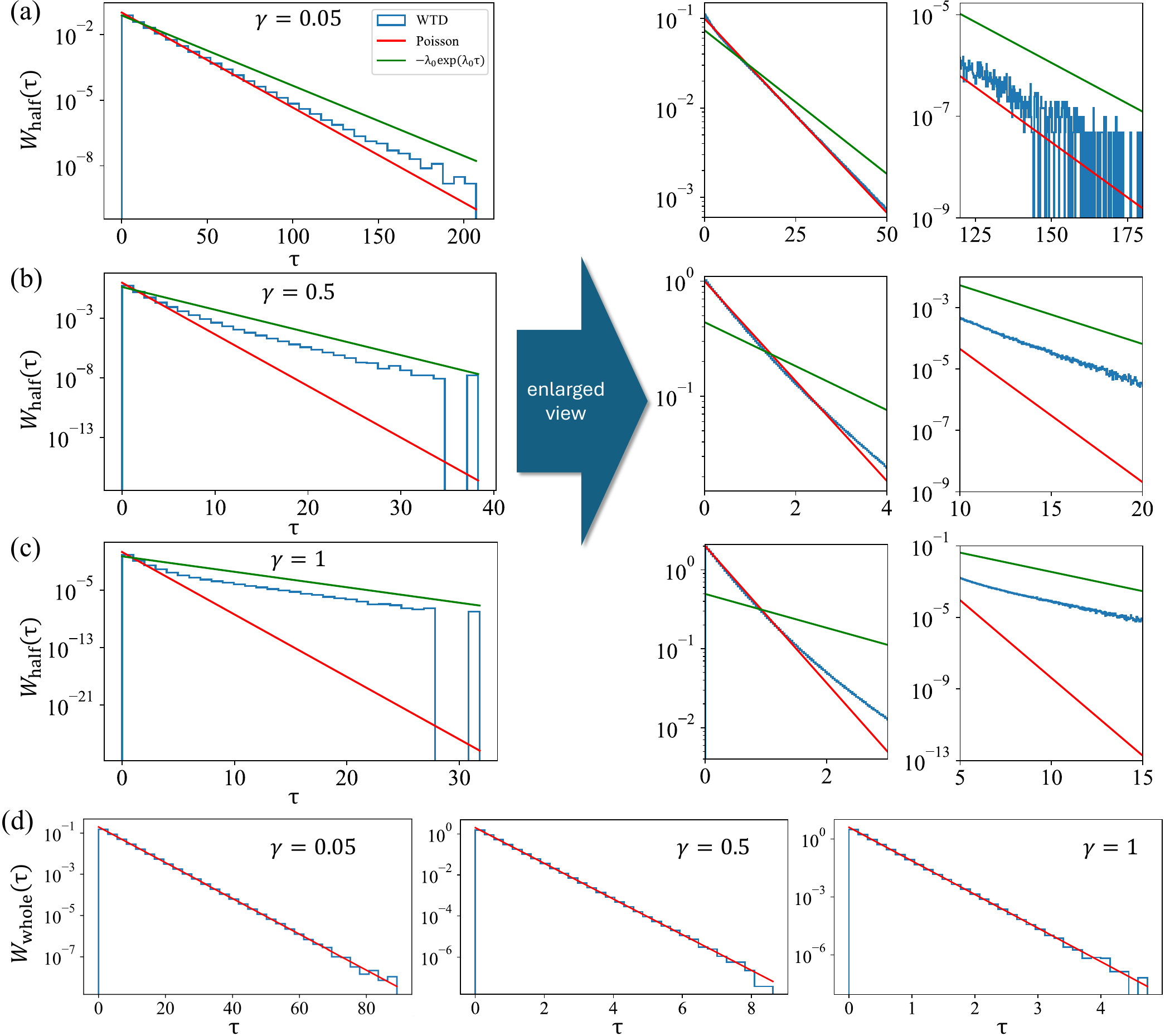}
\caption{Numerical results of the WTD for the Heisenberg model under continuous monitoring with $L=8$ for $\gamma=0.05$ [(a)], $0.5$ [(b)], and $1$ [(c)], demonstrating the emergence of the anomalous tail in the half-chain subsystem characterized by $\lambda_0$. Left panel shows the WTD of the half-chain subsystem (blue histogram). 
The red solid line denotes the Poissonian distribution, while the green solid line corresponds to an exponential distribution characterized by the eigenvalue $\lambda_0$ of the superoperator $\mathscr L_0$ (see text). 
Right panels present enlarged views of the short- and long-time regimes of the half-chain WTD shown in the left panel. 
(d) The WTD of the whole system (blue histogram), which is well described by the Poissonian distribution (red line), for $\gamma=0.05$ (left), $0.5$ (middle), and $1$ (right).
The parameter is set to $t_{\mathrm{ss}}=200$, and the number of trajectories is chosen to be $10^8$.
We note that the drop of the WTD in the blue histogram in the long-time regime is coming from the limitation of the number of simulations.}
\label{fig_wtd}
\end{figure}

\section{Analytical and numerical results for waiting-time distributions}
\label{sec_wtd}
\subsection{Whole system}
Before investigating the WTD of a half-chain subsystem, we show that the WTD of the entire system can be solved analytically. Since the time interval between quantum jumps of the whole system [see Eq.~\eqref{eq_stochastic}] is determined by a uniform one through $t=-2\log R / (\gamma L)$ with the random variable $0\le R\le 1$ \cite{Fuji20, Yamamoto23, Yamamoto25}, the WTD of the total system is trivially given by a Poissonian (i.e., exponential) distribution. This is explicitly demonstrated as follows. First, using Eq.~\eqref{eq_WTDij}, the WTD of the whole system is calculated as
\begin{align}
W_{\mathrm{whole}}(\tau) &= \sum_{i,j=1}^{L}W(\tau,j; i)\notag\\
&=\sum_{i,j=1}^{L}\int_0^\infty \mathrm{Tr}[\mathscr L_j e^{\mathscr L_0 \tau}\mathscr L_i e^{\mathscr L_0 t}\rho_\mathrm{ss}]dt,
\label{eq_WTDtotal}
\end{align}
where the normalization condition is given by
\begin{align}
\int_0^\infty d\tau\; W_{\mathrm{whole}}(\tau)=1.
\end{align}
For the whole system satisfying $M\in[1,L]$, the superoperator $\mathscr L_0$ acts as
\begin{align}
\mathscr L_0(\rho)=-i(H_\mathrm{eff}\rho-\rho H_\mathrm{eff}^\dag),
\end{align}
where the effective non-Hermitian Hamiltonian is defined as
\begin{align}
H_{\mathrm{eff}}=H-\frac{i}{2}\sum_{i=1}^L L_i^\dagger{L}_i=H-\frac{i\gamma L}{4},
\label{eq_Heff}
\end{align}
with the jump operator $L_i=\sqrt \gamma n_i$. This allows us to evaluate the integrand in Eq.~\eqref{eq_WTDtotal} as
\begin{align}
\sum_{i,j=1}^{L}\mathrm{Tr}[\mathscr L_j e^{\mathscr L_0 \tau} \mathscr L_i e^{\mathscr L_0 t}\rho_{\mathrm{ss}}]
&=\gamma^2\sum_{i,j=1}^{L}\mathrm{Tr}[n_j e^{\mathscr L_0 \tau} n_i e^{-\frac{\gamma Lt}{2}} \frac{I}{D_0}]
\displaybreak[2]\notag\\
&=\gamma^2\mathrm{Tr}[(\sum_{j=1}^{L}n_j) e^{\mathscr L_0 \tau} (\sum_{i=1}^{L}n_i) e^{-\frac{\gamma Lt}{2}} \frac{I}{D_0}]
\displaybreak[2]\notag\\
&=\gamma^2\mathrm{Tr}[(\sum_{i=1}^L n_i)^2 e^{\mathscr L_0 \tau}e^{-\frac{\gamma Lt}{2}} \frac{I}{D_0}]
\displaybreak[2]\notag\\
&=\left(\frac{\gamma L}{2}\right)^2 e^{-\frac{\gamma L(t+\tau)}{2}},
\displaybreak[2]
\label{eq_WTDtotalnumerator}
\end{align}
where we note that $\rho_{\mathrm{ss}}=I/D_0$. Here, we have used the fact that, since the time evolution generated by the non-Hermitian Hamiltonian $H_\mathrm{eff}=H-{i\gamma L}/{4}$ does not modify the steady state apart from an overall decay factor, the propagator $e^{\mathscr L_0t}$ acts on $\rho_{\mathrm{ss}}$ to yield $e^{-\frac{\gamma L t}{2}}$ and commutes with the total particle-number operator $\sum_{i=1}^L n_i$.
Finally, substituting Eq.~\eqref{eq_WTDtotalnumerator} into Eq.~\eqref{eq_WTDtotal}, we obtain
\begin{align}
W_{\mathrm{whole}}(\tau)=\frac{\gamma L}{2} e^{-\frac{\gamma L\tau}{2}},
\label{eq_WTDtotalpoisson}
\end{align}
which shows that the WTD of the whole system follows the Poisonian distribution, whose tail decays exponentially with the rate proportional to $L$.

By simulating the continuously monitored dynamics of the Heisenberg Hamiltonian \eqref{eq_HB} using the quantum trajectory method following Eq.~\eqref{eq_stochastic}, we obtain the numerical results of the WTDs of the whole system as shown in Fig.~\ref{fig_wtd}(d), whose parameters are set to $L=8$, and $\gamma=0.05$ (left), $\gamma=0.5$ (center), and $\gamma=1$ (right) (also see the results for larger system sizes in \ref{app_L14}). We find that the numerical data demonstrates that the WTD of the whole system [blue histogram in Fig.~\ref{fig_wtd}(d)] obeys the Poissonian distribution (red line) described by Eq.~\eqref{eq_WTDtotalpoisson}.

\subsection{Half-chain subsystem}
We next study the WTD of the half chain by setting the subsystem to $M \in [1, L/2]$, which leads to
\begin{align}
W_{\mathrm{half}}(\tau) &= \sum_{i,j=1}^{L/2}W(\tau,j; i)\notag\\
&=\sum_{i,j=1}^{L/2}\int_0^\infty \mathrm{Tr}[\mathscr L_j e^{\mathscr L_0 \tau}\mathscr L_i e^{\mathscr L_0 t}\rho_\mathrm{ss}]dt,
\label{eq_WTDhalf_1}
\end{align}
with the normalization condition given by
\begin{align}
\int_0^\infty d\tau\; W_{\mathrm{half}}(\tau)=1.
\end{align}
In this case, the superoperator $\mathscr L_0$ for the half-chain is given by
\begin{align}
\mathscr L_0=\mathscr L - \sum_{i=1}^{L/2}\mathscr L_i,
\end{align}
and the infinite-temperature steady state gives $\sum_{i=1}^{L/2}\mathrm{Tr}[\mathscr L_i \rho_{\mathrm{ss}}]=\gamma L/4$.
We note that a quantity similar to Eq.~\eqref{eq_WTDhalf_1} has appeared in calculating current fluctuations in continuously monitored quantum dynamics \cite{Landi24, Yamamoto25}. However, the main difference between Eq.~\eqref{eq_WTDhalf_1} and current fluctuations lies in the propagator governing the time evolution between two jumps. In the former case, such time evolution is described by the superoperator $\mathscr L_0$, whereas in the latter case, it is governed by the Lindblad superoperator $\mathscr L$. This difference originates from the fact that, in calculating WTDs, the time evolution is constrained such that no additional jump occurs in the subsystem $M$ between two jumps, while no such constraint is present in evaluating current fluctuations.

If the WTD of the half-chain subsystem were to follow a Poissonian distribution, one would expect $W_{\mathrm{half}}(\tau)\overset{?}{=}(\gamma L/4) \exp(-\gamma L\tau/4)$. However, this is not the case, and since $\mathscr L_0$ is a superoperator in calculating WTDs of the half-chain subsystem, it is difficult to obtain an exact closed-form expression as in the case of the whole system described by Eq.~\eqref{eq_WTDtotalpoisson}. We therefore evaluate $W_{\mathrm{half}}(\tau)$ using the spectral decomposition of $\mathcal L_0$. First, the WTD can be rewritten as
\begin{align}
W_{\mathrm{half}}(\tau)
&=\sum_{i,j=1}^{L/2}\int_0^\infty\mathrm{Tr}[\gamma n_j e^{\mathscr L_0 \tau}\mathscr L_i e^{\mathscr L_0 t}\frac{I}{D_0}]dt
\label{eq_wtdhalf}
\end{align}
We note that $\rho_{\mathrm{ss}}=I/D_0$ is the unconditional steady state of the Liouvillian $\mathscr L$. Next, we introduce the vectorization of the density matrix. We again assume that $\mathcal L_0$ is diagonalizable and that eigenvalues are sorted to satisfy Eq.~\eqref{eq_eigenvalue}. The WTD \eqref{eq_wtdhalf} can then be written by using Eq.~\eqref{eq_timeevolution} as
\begin{align}
W_{\mathrm{half}}(\tau)&= \sum_{i,j=1}^{L/2}\sum_{\alpha,\beta=0}^{D_0^2-1}\frac{\gamma}{D_0}\int_0^\infty (n_j| e^{\lambda_\alpha \tau}|\rho_\alpha^R)(\rho_\alpha^L|\mathcal L_i e^{\lambda_\beta t}|\rho_\beta^R)(\rho_\beta^L|I)dt\notag\\
&= -\sum_{i,j=1}^{L/2}\sum_{\alpha,\beta=0}^{D_0^2-1}\frac{\gamma e^{\lambda_\alpha \tau}}{D_0 \lambda_\beta}(n_j|\rho_\alpha^R)(\rho_\alpha^L|\mathcal L_i|\rho_\beta^R)(\rho_\beta^L|I),
\label{eq_wtdhalf2}
\end{align}
where $\mathcal L_i=\gamma n_i\otimes n_i$ after vectorization of the density matrix, and $n_i^*=n_i^T=n_i$ in the Fock basis. 
In Eq.~\eqref{eq_wtdhalf2}, we remark that the mode expansion should be employed in the $D_0$-dimensional half-filling sector, since the total particle number is conserved.
Note that, unlike the case of the Liouvillian, $\lambda_0$ is not the zero eigenvalue, and it gives a finite contribution in Eq.~\eqref{eq_wtdhalf2}. When we focus on the long-time tail of the WTD, the dominant contribution in the limit $\tau\to\infty$ comes from the slowest-decaying mode described by $\lambda_0$.
By combining this fact and the numerical results shown in Fig.~\ref{fig_nojumpop}(c), which shows the deviation of the spectrum of $\lambda_0$ from that of the Poissonian distribution, we find that the half-chain WTD exhibits a long-time tail markedly deviating from that of the Poissonian statistics.
We note that the strong U(1) symmetry of the dynamics with the particle number measurement plays a key role to obtain the measurement-induced phenomena unique to subsystems that are not seen in the whole system.

We can also show that, when $\gamma \tau$ is sufficiently small, the WTD for the half chain exhibits approximately the same exponential decay rate as the Poissonian distribution. To this end, we define $A(t)\equiv e^{\mathscr L_0 t}\rho_{\mathrm{ss}}$ in Eq.~\eqref{eq_wtdhalf}. For sufficiently small $\gamma \tau$, the local operator $n_i$ in the steady state $\rho_{\mathrm{ss}}$ of the Liouvillian can be regarded as approximately conserved, and one may approximate $e^{\mathscr L_0 \tau}\sim e^{\mathscr L \tau}e^{-\sum_{i=1}^{L/2}\mathscr L_i \tau}$. Consequently, Eq.~\eqref{eq_wtdhalf} can be evaluated at short times as
\begin{align}
W_{\mathrm{half}}(\tau)
&\sim\frac{\gamma^2 L}{4}\int_0^\infty dt \mathrm{Tr}[e^{\tilde {\mathscr L}_0 \tau}(I)\sum_{i=1}^{L/2}n_i A(t) n_i]\notag\\
&\sim\frac{\gamma^2 L}{4}\int_0^\infty dt \mathrm{Tr}[e^{-\sum_{i=1}^{L/2}\tilde {\mathscr L}_i \tau} e^{\tilde {\mathscr L} \tau}(I)\sum_{i=1}^{L/2}n_i A(t) n_i]\notag\\
&\sim\frac{\gamma^2 L}{4}e^{-\frac{\gamma L\tau}{4}}\int_0^\infty dt \mathrm{Tr}[\sum_{i=1}^{L/2}n_i A(t) n_i]\notag\\
&\sim\left(\frac{\gamma L}{4}\right)^2 e^{-\frac{\gamma L\tau}{4}}\int_0^\infty dt \mathrm{Tr}[A(t)].
\label{eq_wtdhalf3}
\end{align}
Here, we have used the fact that the identity operator is invariant under the time evolution generated by $\tilde {\mathscr L}$. Thus, the decay rate $\exp(-\gamma L\tau/4)$ coincides with that of the Poissonian distribution for the half-chain subsystem.

Figure~\ref{fig_wtd} shows the numerical results of the half-chain WTD (blue histogram) for $\gamma=0.05$ [Fig.~\ref{fig_wtd}(a)], $0.5$ [Fig.~\ref{fig_wtd}(b)], and $1$ [Fig.~\ref{fig_wtd}(c)]. The calculation is performed by using the quantum trajectory method following Eq.~\eqref{eq_stochastic} for the continuously monitored Heisenberg model with the system size $L=8$ (see \ref{app_L14} for the results with larger system sizes). 
In the algorithm, we focus on the half-chain subsystem $M \in [1, L/2]$ (whole system $M \in [1, L]$) for $W_{\mathrm{half}}(\tau)$ [$W_{\mathrm{whole}}(\tau)$], calculating the time interval between the first and second jumps after $t_{\mathrm{ss}}$ over trajectory realizations.
Figure~\ref{fig_wtd} clearly suggests that the long-time tail of the half-chain WTD becomes heavier than the Poissonian distribution (red line), irrespective of the measurement strength. We further observe that the exponential slope of the anomalous tail of the half-chain WTD is governed by the eigenvalue $\lambda_0$ (green line), and that the time interval dominated by $\lambda_0$ systematically extends as $\gamma$ increases. 
We have also confirmed the similar trend as the measurement strength is further increased and that the anomalous WTD is also obtained for the other values of $J_z$ (\cred{see \ref{app_Jz0p5}}).
Importantly, as $\lambda_0$ is independent of the system size for strong measurement strength [see Fig.~\ref{fig_nojumpop}(c) \cred{and Table \ref{tab_lambda0}}], the anomalous tail of the WTD persists in the thermodynamic limit. Moreover, as shown in the enlarged views of Figs.~\ref{fig_wtd}(a), \ref{fig_wtd}(b), and \ref{fig_wtd}(c), the short-time behavior of the half-chain WTD exhibits the exponential slope close to the Poissonian distribution, indicating the validity of Eq.~\eqref{eq_wtdhalf3} at short times. Therefore, the behavior of the WTD of the half-chain subsystem is drastically different from the one of the total system described by the Poissonian distribution: the exponential slope of the half-chain WTD is characterized by the Poissonian one at short times and crossovers to the anomalous tail described by the eigenvalue $\lambda_0$ of the superoperator $\mathscr L_0$.

\section{Conclusions}
\label{sec_conclusion}
In this work, we have investigated WTDs of quantum jumps in continuously monitored quantum many-body systems, whose steady-state density matrix is the trivial infinite-temperature state. Focusing on a half-chain subsystem, we have uncovered a striking departure from the Poissonian distribution observed in the whole system: the half-chain WTD exhibits the anomalous behavior, which is characterized by $\lambda_0$ of the superoperator $\mathscr L_0$. Furthermore, we have identified a qualitative change in the system-size dependence of $\lambda_0$ as the measurement strength is varied, which suggests that the anomalous behavior of the half-chain WTD persists even in the thermodynamic limit. This highlights the subsystem WTD as a novel diagnostic of monitored quantum many-body systems. The WTD can be extracted directly from the spacetime record of quantum jumps $\{t_i,x_i\}$ without postselection, making it experimentally accessible. For instance, hard-core bosons can be realized using a Tonks-Girardeau gas \cite{Paredes04, Kinoshita04}, and continuous measurements corresponding to jump operators $L_i = n_i$ are expected to be implemented by combining off-resonant probe light \cite{Luschen17, Patil15} with quantum gas microscopy \cite{Bakr09, Sherson10}.
From the numerical results shown in Fig.~\ref{fig_wtd}(c) for $\gamma=1$, we find that the half-chain WTD starts to deviate from the Poissonian distribution around $W_\mathrm{half}(\tau)\sim\mathcal O(10^{-2})$, which means that we can experimentally access the anomalous tail of the half-chain WTD with $\sim 10^2$-$10^3$ samples, which are much less than $10^8$.

The present protocol can also be viewed as a kind of subspace measurement scheme. In stroboscopic subspace measurements of a single-particle problem, it is known that the waiting-time statistics can be dominated by a single eigenvalue of the effective time-evolution operator \cite{Barkai25, Barkai25c}. It is interesting to investigate the relation between those results and ours, latter of which is formulated as a many-body problem (see also Refs.~\cite{Gambassi25} and \cite{Liu24}).

While the long-time tail of the half-chain WTD is governed by the eigenvalue $\lambda_0$ of the superoperator $\mathscr{L}_0$ with the largest real part, the role of the remaining eigenvalues is an open question. In particular, it is interesting to clarify how the eigenvalue $\lambda_1$ with the second-largest real part contributes to the half-chain WTD, and whether it leads to characteristic subleading corrections or crossover behavior at intermediate times.
It is also important to exactly identify the crossover time $\tau^*$, where the half-chain WTD starts to deviate from the Poissonian distribution.
Furthermore, the numerical exact diagonalization results of the spectrum of $\mathscr{L}_0$ suggest the possible emergence of exceptional points in a dissipative many-body system \cite{Heiss12, Minganti19, Luitz19, Takemori24, Takemori25}, where eigenvalues and eigenvectors coalesce. Understanding how such exceptional points influence the WTD is an intriguing direction for future work.
Finally, it is an important open problem to explore the universal properties by investigating how anomalous subsystem WTDs manifest themselves in other interacting many-body models.

\ack{This work was supported by JSPS Program for Forming Japan's Peak Research Universities (J-PEAKS) Grant No.\ JPJS00420230008, JST ERATO Grant No.\ JPMJER2302, and KAKENHI Grants No.\ JP24K16982 and No.\ JP25K17327. This work was also partly funded by Hirose Foundation, Fujikura Foundation, Toyota Riken Scholar Program, and Support Center for Advanced Telecommunications Technology Research. The numerical calculations were carried out in part with the help of QuSpin \cite{Weinberg17}.}

\appendix

\renewcommand{\thesection}{Appendix~\Alph{section}}
\renewcommand{\theequation}{\Alph{section}\arabic{equation}}
\setcounter{equation}{0}

\section{Complete positivity of $e^{\mathscr L_0 t}$}
\label{app_cp}
We prove that $e^{\mathscr L_0 t}$ is a completely positive map for $t\geq0$. To this end, we rewrite the superoperator $\mathscr L_0$ \eqref{eq_nojump} as
\begin{align}
\mathscr L_0 = \mathscr L_{\mathrm{NH}} + \mathscr L_\mathrm{J},
\end{align}
with
\begin{align}
\mathscr L_{\mathrm{NH}}(\rho)=-i(H_\mathrm{eff}\rho-\rho H_\mathrm{eff}^\dag),
\label{eq_LNH1}
\end{align}
and
\begin{align}
\mathscr L_\mathrm{J}(\rho)=\sum_{i\in \bar M} \mathscr L_i (\rho) = \sum_{i\in \bar M} L_i \rho L_i^\dag.
\label{eq_LJ}
\end{align}
Here, $H_\mathrm{eff}$ is given in Eq.~\eqref{eq_Heff} (note that $L_i=\sqrt \gamma n_i$ is not required for the complete positivity of $e^{\mathscr L_0 t}$). We remark that Eq.~\eqref{eq_LJ} is the Kraus representation, and thus $\mathscr L_\mathrm{J}$ is a completely positive map. First, defining that
\begin{align}
\Phi_t^{\mathrm{NH}}(\rho)\equiv e^{-iH_\mathrm{eff}t}\rho e^{t(-iH_\mathrm{eff})^\dag},
\end{align}
we obtain
\begin{align}
\frac{d}{dt}\Phi_t^{\mathrm{NH}}(\rho)=-iH_\mathrm{eff}e^{-iH_\mathrm{eff}t}\rho e^{t(-iH_\mathrm{eff})^\dag} + e^{-iH_\mathrm{eff}t}\rho e^{t(-iH_\mathrm{eff})^\dag}(-iH_\mathrm{eff})^\dag=\mathscr L_{\mathrm{NH}}(\Phi_t^{\mathrm{NH}}(\rho)),
\end{align}
which gives
\begin{align}
e^{\mathscr L_{\mathrm{NH}} t}(\rho)= \Phi_t^{\mathrm{NH}}(\rho)=e^{-iH_\mathrm{eff}t}\rho e^{t(-iH_\mathrm{eff})^\dag},
\label{eq_LNH}
\end{align}
where we have used $\Phi_{t=0}^{\mathrm{NH}}(\rho)=\rho$. Equation \eqref{eq_LNH} is the Kraus representation, and thus $e^{\mathscr L_{\mathrm{NH}} t}$ is a completely positive map. In addition, since Eq.~\eqref{eq_LJ} is the Kraus representation, we find that $e^{\mathscr L_\mathrm{J}t}$ is immediately a completely positive map. Finally, by using these facts in the Trotter product formula, given by
\begin{align}
e^{\mathscr L_0 t} = e^{(\mathscr L_{\mathrm{NH}} + \mathscr L_\mathrm{J})t}=\lim_{n\to\infty}\left(e^{\frac{\mathscr L_{\mathrm{NH}}t}{n}}  e^{\frac{\mathscr L_\mathrm{J}t}{n}}\right)^n,
\end{align}
we conclude that $e^{\mathscr L_0 t}$ is a completely positive map.

\begin{figure}[tb]
\centering
\includegraphics[width=0.9\textwidth]{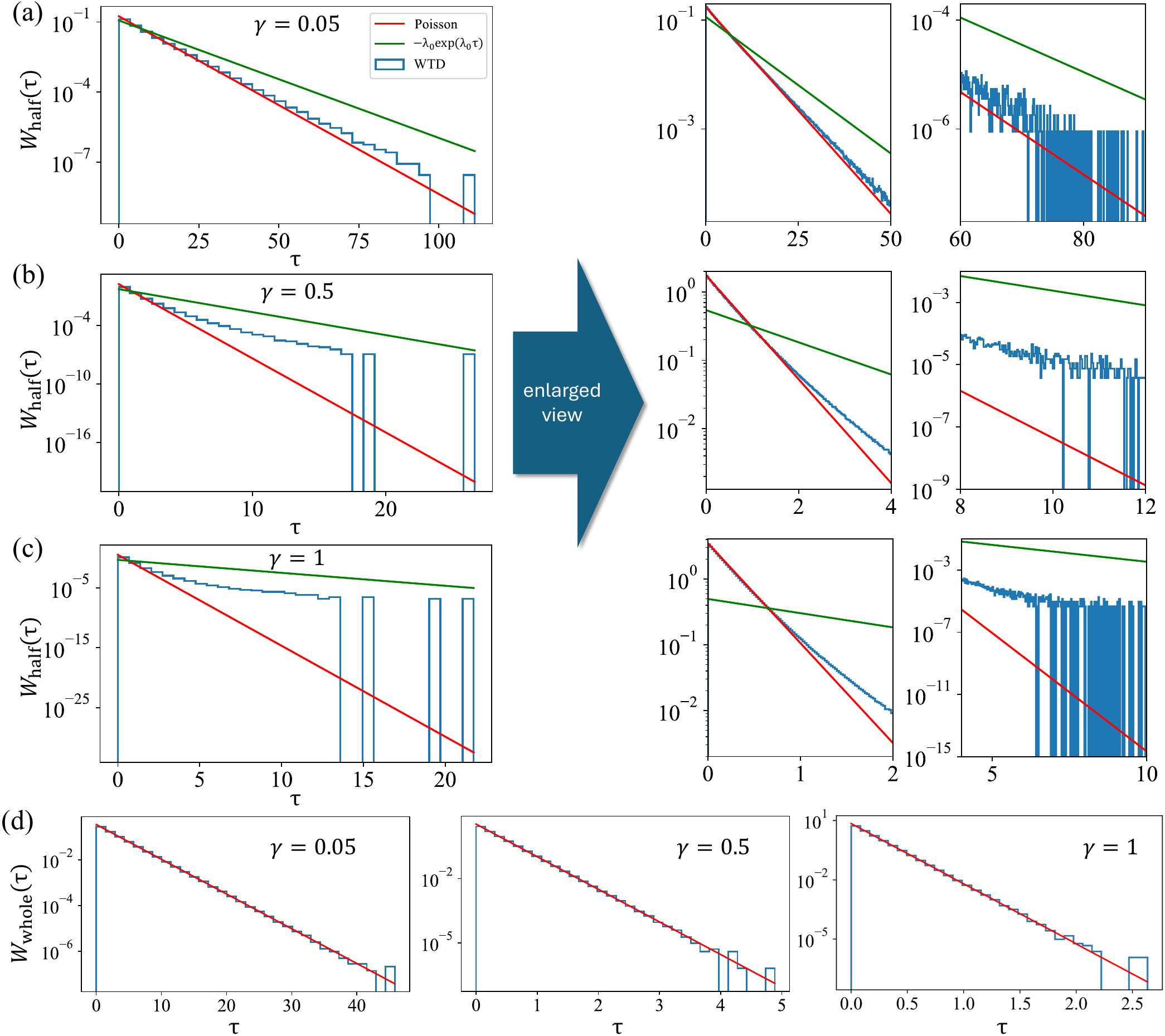}
\caption{Numerical results of the WTD for the Heisenberg model under continuous monitoring with $L=14$ for $\gamma=0.05$ [(a)], $0.5$ [(b)], and $1$ [(c)], demonstrating the emergence of the anomalous tail in the half-chain subsystem characterized by $\lambda_0$. Left panel shows the WTD of the half-chain subsystem (blue histogram). 
The red solid line denotes the Poissonian distribution, while the green solid line corresponds to an exponential distribution characterized by the eigenvalue $\lambda_0$ of the superoperator $\mathscr L_0$ (see text). 
Right panels present enlarged views of the short- and long-time regimes of the half-chain WTD shown in the left panel.
(d) The WTD of the whole system (blue histogram), which is well described by the Poissonian distribution (red line), for $\gamma=0.05$ (left), $0.5$ (middle), and $1$ (right).
The parameter is set to $t_{\mathrm{ss}}=200$, and the number of trajectories is chosen to be $10^7$.
We note that the drop of the WTD in the blue histogram in the long-time regime is coming from the limitation of the number of simulations.}
\label{fig_wtdl14}
\end{figure}

\section{Numerical results for waiting-time distributions in large system sizes}
\label{app_L14}
To confirm that the numerical results obtained in Fig.~\ref{fig_wtd} correctly capture the ones in large system sizes, we further perform numerical calculations of the half-chain WTD for $L=14$. As shown in Fig.~\ref{fig_wtdl14} for $\gamma=0.05$ [(a)], $0.5$ [(b)], and $1$ [(c)], the long-time tail of the half-chain WTD (blue histogram) becomes heavier than the Poissonian distribution (red line), and the exponential slope of the anomalous tail is governed by $\lambda_0$ (green line). Though the time interval dominated by $\lambda_0$ is narrow for weak measurement strength, it gradually extends as $\gamma$ is increased. Also, the short-time behavior of the half-chain WTD exhibits the exponential slope close to the Poissonian distribution as shown in the enlarged views of Figs.~\ref{fig_wtdl14}(a), \ref{fig_wtdl14}(b), and \ref{fig_wtdl14}(c). We also see that the Poissonian distribution of the WTD of the entire system obeys the standard Poissonian distribution as shown in Fig.~\ref{fig_wtdl14}(d). Therefore, the characteristic behavior of the WTD obtained in Fig.~\ref{fig_wtd} persists even in large system sizes. We remark that large system sizes require a large number of trajectories to capture the long-time behavior of the WTD, making numerical simulations harder.

\begin{figure}[tb]
\centering
\includegraphics[width=0.9\textwidth]{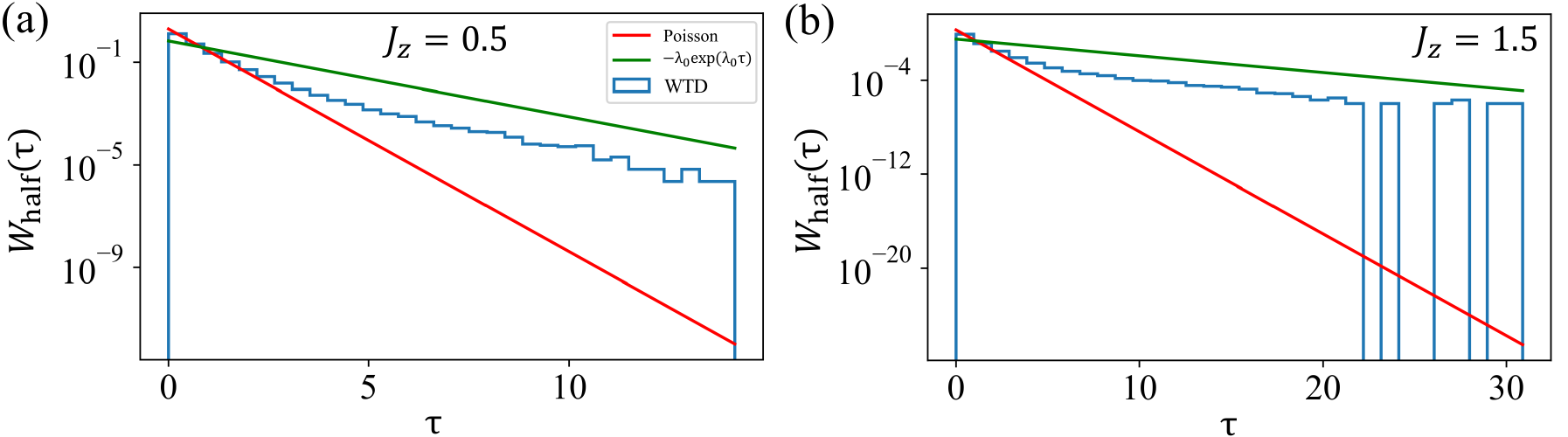}
\caption{
\cred{Numerical results of the WTD for the XXZ model under continuous monitoring with [(a)] $J_z=0.5$ and [(b)] $J_z=1.5$, demonstrating the emergence of the anomalous tail in the half-chain subsystem (blue histogram) characterized by $\lambda_0$.
The red solid line denotes the Poissonian distribution, while the green solid line corresponds to an exponential distribution characterized by the eigenvalue $\lambda_0$ of the superoperator $\mathscr L_0$. 
The parameters are set to $L=8$, $\gamma=1$, and $t_{\mathrm{ss}}=200$, and the number of trajectories is chosen to be $10^6$.
We note that the drop of the WTD in the blue histogram in the long-time regime is coming from the limitation of the number of simulations.}
}
\label{fig_wtdotherJz}
\end{figure}

\section{Numerical results for waiting-time distributions in XXZ models}
\label{app_Jz0p5}
\cred{To confirm that the anomalous tail of the WTD appears irrespective of the interaction strength, we further perform numerical calculations of the half-chain WTD in XXZ models.
As shown in Fig.~\ref{fig_wtdotherJz} for $J_z=0.5$ [(a)] and $1.5$ [(b)], we see that the long-time tail of the half-chain WTD (blue histogram) becomes heavier than the Poissonian distribution (red line), and the exponential slope of the anomalous tail is governed by $\lambda_0$ (green line).  Therefore, the anomalous WTD persists irrespective of the interaction strengths.}


\providecommand{\newblock}{}

\end{document}